\documentclass{aa}
\usepackage{graphicx}
\begin{document}
\newcommand\beq{\begin{equation}}
\newcommand\eeq{\end{equation}}
\newcommand\mum{\mu{m}}
\def\DeltaRA{|\Delta{\rm RA|}}
\def\DeltaDec{|\Delta{\rm DEC|}}

{\title{Age Estimations of M31 Globular Clusters from Their
Spectral Energy Distributions}}

\author{Jun Ma\inst{1}, Xu Zhou\inst{1}, David Burstein\inst{2},
Yanbin Yang\inst{1}, Zhou Fan\inst{1}, Jiansheng Chen\inst{1},
Zhaoji Jiang\inst{1}, Zhenyu Wu\inst{1}, Jianghua Wu\inst{1} and
Tianmen Zhang\inst{1}}

\institute{National Astronomical Observatories, Chinese Academy of
Sciences, Beijing, 100012, P. R. China \and Department of Physics
and Astronomy, Box 871504, Arizona State University, Tempe, AZ
85287--1504}

\offprints{Jun Ma, \\
\email{majun@vega.bac.pku.edu.cn}}

\date{Received / Accepted}

\abstract{This paper presents accurate spectral energy
distributions (SEDs) of 16 M31 globular clusters (GCs) confirmed
by spectroscopy and/or high spatial-resolution imaging, as well as
30 M31 globular cluster candidates detected by Mochejska et al. (1998).
Most of these candidates have $m_V > 18$, deeper than previous
searches, and these candidates have not yet been confirmed to be
globular clusters. The SEDs of these clusters and candidates are
obtained as part of the BATC Multicolor Survey of the Sky, in
which the spectrophotometrically-calibrated CCD images of M31 in
13 intermediate-band filters from 4000 to 10000{\AA} were
observed. These filters are specifically designed to exclude most
of the bright and variable night-sky emission lines including the
OH forest. In comparison to the SEDs of true GCs, we find that
some of the candidate objects are not GCs in M31. SED fits show
that theoretical simple stellar population (SSP) models can fit
the true GCs very well. We estimate the ages of these GCs by
comparing with SSP models. We find that, the M31 clusters range in
age from a few ten Myr to a few Gyr old, as well as old GCs,
confirming the conclusion that has been found by Barmby et al. (2000),
Williams \& Hodge (2001), Beasley et al. (2004), Burstein et al.
(2004) and Puzia et al. (2005) in their investigations of the
SEDs of M31 globular clusters.
}

\titlerunning{Globular Clusters and Candidates in M31}
\authorrunning{J. Ma et al.}
\maketitle

\section{Introduction}

Galactic globular clusters are among the oldest stellar
objects in the universe, provide vitally important information
regarding the minimum age of the universe and the early formation
history of our Galaxy. Studying the integrated properties of
extragalactic globular clusters can help us understand the
evolutionary history of distant galaxies (cf. \cite{bur04}).
Globular clusters (GCs) are bright, easily identifiable stellar
populations with homogeneous abundances and ages.
M31 is an ideal target for studying GCs, since it comprises the
largest and nearest sample of GCs, which is more than all GCs
combined in the other Local Group members (\cite{battis87};
\cite{raci91}; \cite{harris91}; \cite{fusi1993}).

Many GC searches in M31 have been conducted. The first catalog of
M31 GC was presented by Hubble (1932), who discovered 140 GCs with
$m_{pg}\leq 18$ mag. Then, a number of catalogs of GC candidates
were published. Vete\u{s}nik (1962a) compiled the first major
catalog, containing about 300 GC candidates identified by the
previous works (e.g., \cite{hubble32}; \cite{seynas45};
\cite{hilt58}; \cite{mayegg53}; \cite{kronmay60}).  Later, several
other major catalogs of M31 GC candidates were compiled by Sargent
et al. (1977), Crampton et al. (1985) and by the Bologna Group
(\cite{battis80,battis87,battis93}). Although these catalogs may
be fairly complete down to $V=18$ ($M_v \sim -6.5$)
(\cite{fusi1993}), recent works have searched for fainter GCs in
M31 (e.g., \cite{moche98}; \cite{bh01}).

These latter samples provide a good database for studies of M31
GCs, including their luminosity function (e.g., \cite{auri92};
\cite{moche98}; \cite{barmby01}), reddening and intrinsic color
(e.g., \cite{vete62b}; \cite{bage77}; \cite{iyeri85}; \cite{bh00};
\cite{barmby02b}), metallicities (e.g., \cite{vand69};
\cite{ash93}; \cite{bh00}; \cite{perr02}), structure
parameters (\cite{barmby02a}) and comparisons with the Galactic
GCs and M33 GCs (e.g., \cite{hilt60}; \cite{fpc80}; \cite{rhh92};
\cite{moche98}). Near infrared photometry of M31 GCs has been done
as well (e.g., \cite{fpc80}; \cite{Sitko}; \cite{Bonolia,Bonolib};
\cite{Cohen}; \cite{bh00}), with Galleti et al. (2004) using 2MASS
data for 693 GC candidates.  In a previous paper from our group,
Jiang et al. (2003) obtained CCD multicolor photometry for 172 M31
GCs and candidates in 13 intermediate-band filters spanning
wavelengths from 4000 to 10000{\AA}.

Barmby et al. (2000); Beasley et al. (2004); Burstein et al.
(2004) and Puzia et al. (2005), using SEDs, have found a set of
young to intermediate-age GCs in M31, as well as the ``usual''
complement of GCs as old as Galactic GCs. Jiang et al. (2003)
estimated ages for their M31 GCs by comparing their SEDs to those
of theoretical simple stellar population (SSP) models, finding at
least 8 M31 GCs and candidates younger than 1 Gyr. Recently, Fusi
Pecci et al. (2005) found a population of 67 massive blue clusters
in the disk of M31, which they interpret as globular clusters,
with ages less than $\sim2~\rm{Gyr}$.

M31 was observed as part of galaxy calibration program of the
Beijing-Arizona-Taiwan-Connecticut (BATC) Multicolor Sky Survey
(e.g., \cite{fan96}; \cite{zheng99}), which has a custom-designed
set of 15 intermediate-band filters to do spectrophotometry for
preselected 1 deg$^{2}$ regions of the northern sky. In this
paper, we present the SEDs for 15 M31 GCs confirmed by
spectroscopy and/or high spatial-resolution imaging and 30 M31 GC
candidates detected by Mochejska et al. (1998) that lie within the
BATC field of view. Details pertaining to the observations and
data reduction are given in \S~2. Analysis of the M31 GC SEDs is
reported in \S~3. A summary is given in \S~4.

\section{Observations and Data Reduction}

\subsection{Sample of GCs and GC Candidates}

The sample GCs and GC candidates in this paper is from Mochejska
et al. (1998), who found 105 GC candidates using the data
collected in the DIRECT project (\cite{Kaluzny98};
\cite{Stanek98}). DIRECT observations are done with the 1.2 m
telescope at the F.L. Whipple Observatory. The
pixel scale for the DIRECT observations is
$0\arcsec{\mbox{}\hspace{-0.15cm}.} 3$. Mochejska et al. (1998)
present photometry for these candidates using
standard Johnson-Cousins $VI$ filters. We were able to find 65
objects in the Mochejska et al. (1998) sample in the BATC CCD
field. As 19 candidates either have bright stars nearby or have
too much M31 background, and one candidate is at the edge of the
BATC CCD images in most filter bands, we do not present
spectrophotometry for them (M004, M008, M014, M019, M031, M034,
M039, M040, M041, M042, M044, M046, M050, M054, M055, M060, M062,
M068, M070 and M072). Mxxx means that this candidate is from
Mochejska et al. (1998) in order to be consistent with the
nomenclature of Galleti et al. (2004). Of the remaining 45 GC
candidates, 15 are confirmed GCs that have also been detected by
other authors (see details from \cite{gall04}). The SEDs of these
GCs were also given by Jiang et al. (2003).

\subsection{Observations and Data Reduction}

The BATC survey uses a Ford Aerospace $2048\times 2048$ CCD camera
with 15 $\mu\rm{m}$ pixel size, mounted at the focus of the
0.6/0.9m f/3 Schmidt telescope located at the Xinglong Station of
the National Astronomical Observatories of China. The typical
seeing of the Xinglong station is $2\arcsec$. The multicolor
BATC filter system was specifically designed to avoid
contamination from the brightest and most variable night
sky emission lines. Our 15 intermediate-band filters cover the
wavelength range from  3300{\AA} to 1$\mu$.
Spectrophotometric calibrations of these images are made using
observations of four $F$ sub-dwarfs, HD~19445, HD~84937,
BD~${+26^{\circ}2606}$, and BD~${+17^{\circ}4708}$, all taken from
Oke \& Gunn (1983). Hence, our magnitudes are defined in a way
similar to the spectrophotometric AB magnitude system (i.e, the
Oke \& Gunn $\tilde{f_{\nu}}$ monochromatic system). BATC
magnitudes are defined on the AB magnitude system as

\begin{equation}
m_{\rm batc}=-2.5{\rm log}\tilde{F_{\nu}}-48.60,
\end{equation}

\noindent where $\tilde{F_{\nu}}$ is the appropriately averaged
monochromatic flux in unit of erg s$^{-1}$ cm$^{-2}$ Hz$^{-1}$ at
the effective wavelength of the specific passband. In the BATC
system (\cite{yan00}), $\tilde{F_{\nu}}$ is defined as

\begin{equation}
\tilde{F_{\nu}}=\frac{\int{d} ({\rm log}\nu)f_{\nu}r_{\nu}}
{\int{d} ({\rm log}\nu)r_{\nu}},
\end{equation}

\noindent which links the magnitude to the number of photons
detected by the CCD rather than to the input flux (\cite{fuku96}).
In Equation (2), $r_{\nu}$ is the system's response, $f_{\nu}$ is
the SED of the source.

37 hours of images of the BATC M31 field were obtained in
13 intermediate-band filters (excluding the two bluest filters)
from September 15, 1995 to December 16, 1999. Bias subtraction and
flat-fielding with dome flats were done with the BATC automatic
data reduction software, PIPELINE I, developed for the BATC
Multicolor Sky Survey (\cite{fan96}; \cite{zheng99}). The dome
flat-field images were taken by using a diffuser plate in front of
the correcting plate of the Schmidt telescope, a flatfielding
technique which has been verified with the photometry we have done
on other galaxies and spectrophotometric observations (e.g.,
\cite{fan96}; \cite{zheng99}; \cite{wu02}; \cite{yan00};
\cite{zhou01}). Spectrophotometric calibration of the M31 images
using the Oke-Gunn standard stars is done during photometric
nights (see details from \cite{yan00}; \cite{zhou01}).

Using the images of the standard stars observed on photometric
nights, we iteratively obtain atmospheric extinction curves
and variation of these extinction coefficients with time of night
(cf. \cite{yan00}; \cite{zhou01}):

\begin{equation}
m_{\rm batc}=m_{\rm inst}+[K+\Delta K(UT)]X+C,
\end{equation}

\noindent where $X$ is the air mass and $K(UT)$ is the
time-dependent extinction term. The instrumental magnitudes
($m_{\rm inst}$) of select bright, isolated and unsaturated
stars on the M31 field images on photometric nights can be readily
transformed to the BATC AB magnitude system ($m_{\rm batc}$). The
calibrated magnitudes of these stars are then used as secondary
standards to uniformly combine images from calibrated nights to
those taken during non-photometric weather. Table~1 lists the
parameters of the BATC filters and the statistics of observations.
Column 6 of Table 1 gives the scatter, in magnitudes, for the
photometric observations of the four primary standard stars in
each filter.

\subsection{Integrated Photometry}

For each M31 GC candidate, the PHOT routine in DAOPHOT
(\cite{stet87}) is used to obtain magnitudes. To avoid
contamination from nearby objects, we adopt an aperture of
$10\arcsec{\mbox{}\hspace{-0.15cm}.}2$ corresponding to a diameter
of 6 pixels on the Ford CCD. Inner and outer radii for background
determination are taken at 8 to 13 pixels from the center of the GCs.
Given the small aperture use for the GC observations,
aperture corrections are determined as follows:  We use the
isolated stars to determine the magnitude difference beween
photometric radii of 6 pixels and the full magnitude of these
stars in each of the 13 BATC filters. The spectral energy
distributions (SEDs) for 45 GC candidates are then corrected for
this difference in each filter, and these values are
given in Table 2. Column 2 to Column
14 give the magnitudes of the 13 BATC passbands observed.  The
second line for each object gives the $1-\sigma$ errors in
magnitudes for the corresponding passband. The errors for each
filter are given by DAOPHOT. The magnitudes of some objects in
some BATC filters could not be obtained owing to low
signal-to-noise ratio in these filters, or they lie at the edge of
the BATC CCD images. In this table, Mxxx means that this candidate
is from Mochejska et al. (1998) in order to be consistent with the
nomenclature of Galleti et al. (2004).

\begin{figure*}
\begin{center}
\includegraphics[angle=-90,width=160mm]{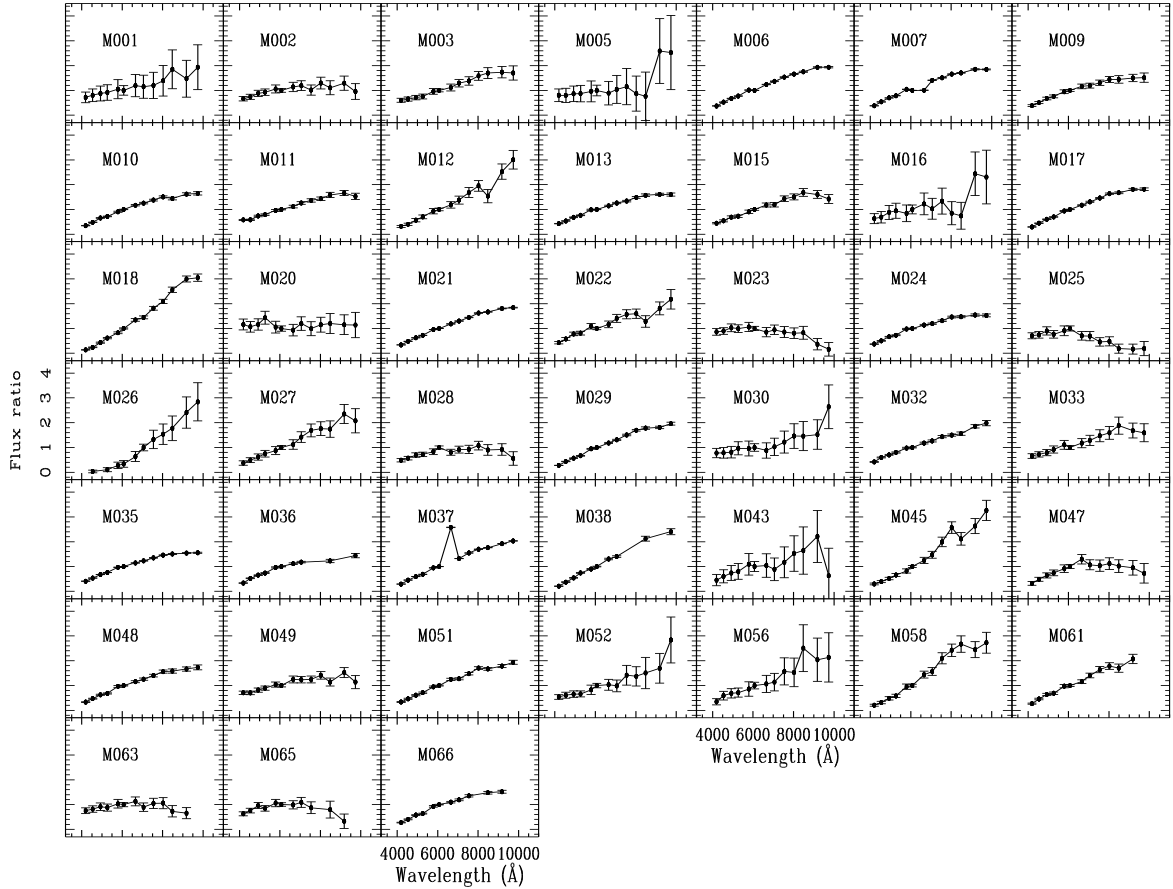}
\caption{Spectral energy distributions for 45 M31
GCs and GC candidates.}
\label{fig:one}
\end{center}
\end{figure*}

\subsection{SEDs of Some Objects Are Found not to be GCs}

Figure 1 shows the SEDs for 15 M31 confirmed GCs and 30 GC
candidates in the 13 BATC filters. For convenience, we calculate
the ratio of flux relative to the filter BATC08
($\lambda=6075${\AA}) (Since M026 has very strong emissions in red
filters, we use the ratio of flux relative to the filter BATC10
($\lambda=7010${\AA}) in order that it can be plotted in Fig. 1.).
As indicated in Section 2.1, there are 15 known GCs:  M006, M007,
M010, M011, M013, M015, M017, M021, M024, M029, M032, M035, M037,
M048, and M051 (see details from \cite{gall04}). From the SEDs of
these GCs, we can see that, except for M037, the SEDs of the other
GCs do not vary steeply from one BATC filter band to another. In
contrast, GC candidates M005, M016, and M043 are likely not M31
GCs. GC candidate M026 is very red and worthy of future study.

\section{Stellar Population Model and Confirm GCs}

\subsection{The BC03 Models and SED fits}

Bruzual \& Charlot (2003) (hereafter BC03) provide spectral
photometric properties for a wide range of stellar metallicities.
BC03 provide 26 SSP models (high resolution and low resolution)
using the Padova 1994 evolutionary tracks.  13 of these SSP models
are computed using the Chabrier (2003) IMF with lower and upper
mass cutoffs $m_{L}=0.1 M_{\odot}$ and $m_{U}=100 M_{\odot}$. The
other other 13 are computed using the Salpeter (1955) IMF with the
same mass cutoffs. In addition, BC03 provides 26 SSP models using
the Padova 2000 evolutionary tracks. However, as Bruzual \&
Charlot (2003) pointed out that, the Padova 2000 models, which
include more recent input physical parameters than the Padova 1994
models, tend to produce worse agreement with observed galaxy
colors. These SSP models contains 221 spectra describing the
spectral evolution of an SSP from 0 to 20Gyr. The evolving spectra
include the contribution of the stellar component in the range
from 91\AA~~to $160\mu m$. In this paper, we adopted the SSP
models (high resolution) computing using the Padova 1994
evolutionary tracks and Salpeter (1955) IMF.

To proceed with the comparisons, we first convolve the SEDs of
BC03 models with the BATC filter profiles to obtain the optical
and near-infrared integrated luminosities. The integrated
luminosities $L_{\lambda_i}(t,Z)$ of the $i$th BATC filter can be
calculated as

\begin{equation}
L_{\lambda_i}(t,Z) =\frac{\int
F_{\lambda}(t,Z)\varphi_i(\lambda)d\lambda} {\int
\varphi_i(\lambda)d\lambda},
\end{equation}

\noindent where $F_{\lambda}(t,Z)$ is the SEDs at age $t$ in
metallicity $Z$ model, $\varphi_i(\lambda)$ is the response
functions of the $i$th filter of the BATC filter system ($i=3, 4,
\cdot\cdot\cdot, 15$), respectively. All integrated colors of BC03
models are calculated relative to the BATC filter BATC08
($\lambda=6075${\AA}):

\begin{equation}
\label{color}
C_{\lambda_i}(t,Z)={L_{\lambda_i}(t,Z)}/{L_{6075}(t,Z)}.
\end{equation}

\noindent From this equation, we can obtain model
intermediate-band colors. We use the $\chi2$ test to examine
which SED families in BC03 SSP models are most compatible with
those of the observed GCs.

\begin{equation}
\chi2=\sum_{i=3}^{15}{\frac{[C_{\lambda_i}^{\rm
intr}(n)-C_{\lambda_i}^{\rm ssp}(t, Z)]^2}{\sigma_{i}^{2}}},
\end{equation}

\noindent where $C_{\lambda_i}^{\rm ssp}(t, Z)$ represents the
integrated color in the $i$th filter of a SSP at age $t$ in a
metallicity $Z$ model. $C_{\lambda_i}^{\rm intr}(n)$ is the
intrinsic integrated color for a GC, and

\begin{equation}
\sigma^{2}=\sigma_{\rm obs}^{2}+\sigma_{\rm mod}^{2},
\end{equation}

\noindent $\sigma_{\rm obs}^{2}$ is the observational error, and
$\sigma_{\rm mod}^{2}$ is the uncertainty from the model itself.
Charlot et al. (1996) estimate the model uncertainty, $\sigma_{\rm
mod}^{2}$, by comparing the colors obtained from different stellar
evolutionary tracks and spectral libraries. As Wu et al. (2005)
did, we also adopted $\sigma_{\rm mod}^{2}=0.05$ in this paper.
BC03 SSP models using in this paper include six initial
metallicities: 0.0001, 0.0004, 0.004, 0.008, 0.02, and 0.05.
Spectra for other metallicities can be obtained by linear
interpolation between these six base spectra.

\subsection{Reddening Correction of Sample GCs and Candidates}

The observed colors of sample clusters or candidates are affected
by two sources of reddening: the foreground extinction in the
Milky Way and internal reddening in M31.

The Galactic reddening in the direction of M31 was estimated by
many authors (e.g., \cite{vand69}; \cite{McRa69}; \cite{fpc80}),
and similar values of the foreground color excess, $E(B-V)$, were
determined, such as $E(B-V)=0.08$ given by van den Bergh (1969),
0.11 given by McClure \& Racine (1969), and 0.08 given by Frogel
et al. (1980). The reddenings of 12 GCs are kindly given to us by
P. Barmby. For the other objects, we accept the mean reddening,
$E(B-V)=0.11$, as Fusi Pecci et al. (2005) did. The values of
extinction coefficient $R_{\lambda}$ are obtained by interpolating
the interstellar extinction curve of Cardelli et al. (1989).

\subsection{SED Fits between SSP Models and GCs}

In this paper, there are 15 known GCs confirmed by spectroscopy
and/or high spatial-resolution imaging (see details from
\cite{gall04}). We indicate that, M037 (= Bo225 from
\cite{battis87}), was confirmed to be a GC by spectroscopy
and high spatial-resolution imaging, has an emission line in the
BATC09 filter band (the central wavelength is 6710 \AA) and its
SEDs cannot be fit by SSP models. (It is likely that this
emission line comes from the disk of M31, not from the GC itself.)

M31 GCs generally have a metal abundance, [Fe/H], lower than 0.3
and higher than $-2.0$ (\cite{bh00}), so, we only use models of
metallicities between 0.0004 and 0.02 of BC03. Table 3 lists the
results of SED fits. ``Bo'' means that this GC is also detected by
Battistini et al. (1987). Age and metallicity mean that SEDs
of GCs can best be fit by SSPs at this age in this
metallicity model. Figure 2 plots the SED fits. In this figure,
the open circle represents the intrinsic integrated color of the
sample GCs, and the thick line represents the best fit of the
integrated color of a BC03 model. Figure 2 shows that the SEDs of
these GCs can be fit very well by BC03 models.

Extragalactic globular cluster ages are inferred
from composite colors and/or spectroscopy. In the case of
extragalactic GCs, young ages can be interpreted by a strong
Balmer line spectrum, and/or a significant Balmer jump. The above
results can tell us that, M31 includes a population of
intermediate-age GCs with ages of a few Gyr, in agreement with the
conclusions of Beasley et al. (2004), Burstein et al. (2004) and
Puzia et al. (2005) in their investigations of the SEDs of M31
globular clusters.

Figure 3 shows contours of $\Delta \chi2=\chi2-\chi2_{min}$ in
the age-metallicity plane for a few globular clusters randomly
selected. It is true that the age-metallicity degeneracy exists,
so it is difficult to estimate the age and metallicity
simultaneously for individual clusters. However, for statistical
purposes, we worked out the best fit of age and metallicity for
the sample clusters  from ages of 1 to 11 Gyr and metallicities
 of 0.02 to 1$Z_\odot$.

\begin{figure}
\resizebox{\hsize}{!}{\rotatebox{-90}{\includegraphics{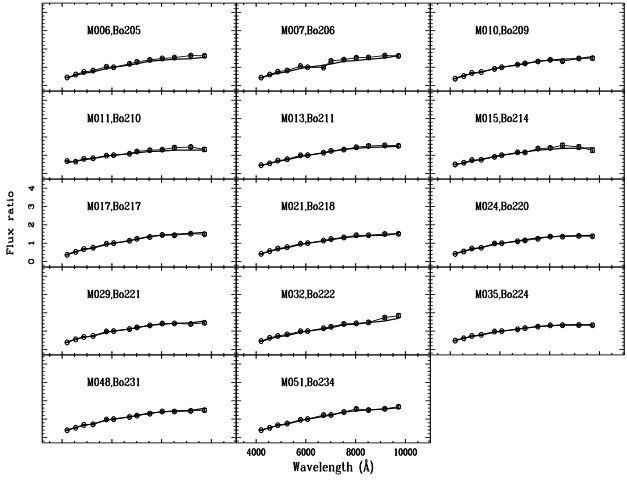}}}
\vspace{0.0cm} \caption{Graphs of the best-fit integrated colors
of SSP models plotted on top of the intrinsic integrated colors
for 14 GCs. Open circles with error bars represent the intrinsic
integrated color of a GC, the thick line represents the best fit
of the integrated color of a SSP model. The $Y$-axis is the ratio
of the flux in each filter to the flux in filter BATC08.}
\label{fig:two}
\end{figure}

\subsection{Comparison of Results}

The observed colors of sample clusters or candidates in this paper are
affected by two sources of reddening: the foreground extinction in
the Milky Way and internal reddening in M31. There are few
determinations of reddening for each individual cluster in M31.
Barmby et al. (2000) determined reddening values for M31 GCs
using the correlations between optical and infrared colors
and metallicity by defining various ``reddening-free'' parameters
with a large database of multicolor photometry. For showing
uncertainty of the derived ages and metallicities due to the
uncertain reddening, we re-estimate the ages and metallicities for
the clusters in Figure 3 as examples using the mean reddening
$E(B-V)=0.11$. The results are listed in Table 4. From this table,
we can see that the uncertainty of the derived ages and
metallicities caused by the uncertain reddening are smaller than 3
Gyr and 0.6 dex, respectively.

\begin{figure}
\resizebox{\hsize}{!}{\rotatebox{-90}{\includegraphics{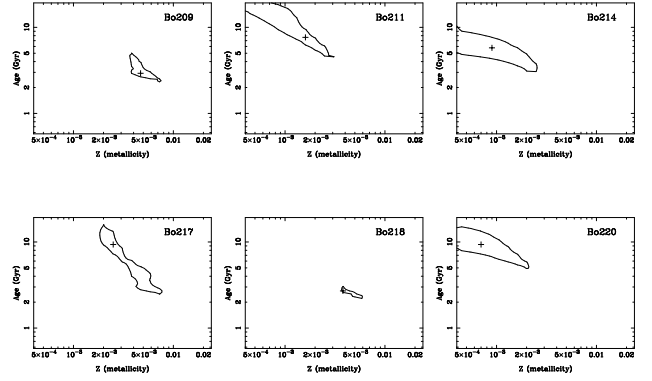}}}
\vspace{0.0cm} \caption{Contours of $\Delta \chi2=2.30$ for a few
M31 globular clusters.} \label{fig:three}
\end{figure}

\subsection{SED fits of between SSP Model and GC Candidates}

\begin{figure}
\resizebox{\hsize}{!}{\rotatebox{-90}{\includegraphics{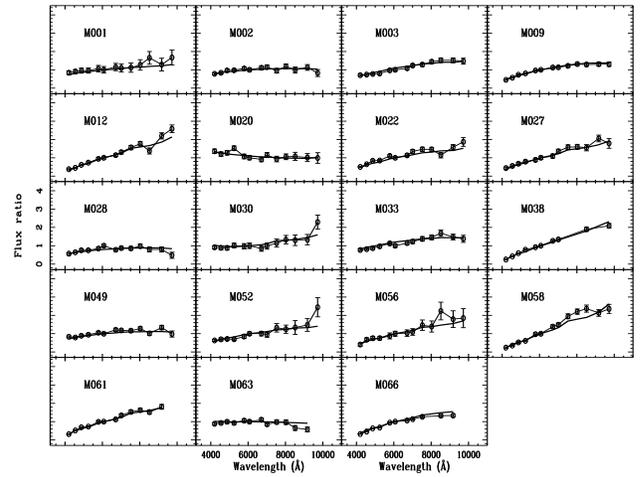}}}
\vspace{0.0cm} \caption{Graphs of the best-fit integrated colors
of SSP models plotted on top of the intrinsic integrated colors
for 19 confirmed GCs. Open circles with error bars represent the
intrinsic integrated color of an object, the thick line represents
the best fit of the integrated color of a SSP model. The $Y$-axis
is the ratio of the flux in each filter to the flux in filter
BATC08.} \label{fig:three}
\end{figure}

We have demonstrated that the SEDs of GCs can be fit very well by
SSP models. From Table 3, we find that the fitting-parameter
$\chi2_{min} \le 2.64$ for all these GCs. We take this criterion
as indicating a good fit between the SSPs and the GC candidates.
The results of fits between SSPs and GC candidates are listed in
Table 5. If the fitting-parameter $\chi2_{min} > 2.64$, we can
say that there is not an appropriate SSP to fit this candidate.
From Table 5, we can see that there is 7 very young GCs, younger
than 1 Gyr. Williams \& Hodge (2001) obtained deep
$Hubble~~Space~~Telescope$ (HST) photometry of individual stars
and color-magnitude diagram (CMD) for four young M31 disk clusters,
subsequently identified as M31 GCs, giving ages in the range
60-160 Myr. Beasley et al. (2004) also found an M31 GC with an
age of 10-30 Myr.

Figure 4 gives the SED fits for confirmed GCs: open circles
represent intrinsic integrated colors of our sample GCs, and
thick lines represent the best fit of the integrated
color of a SSP model. Cohen et al. (2005) present observations
of 6 objects in M31 that are alleged to be young globular
clusters, and find that the four youngest of these objects are
asterisms. So, we should keep in mind that the younger globular
cluster candidates in this paper may be dense associations of
younger stars in M31.

Combining Tables 3 and 5, we plot the location of the clusters
ontop of M31 image in Figure 5.

\begin{figure}
\resizebox{\hsize}{!}{\rotatebox{0}{\includegraphics{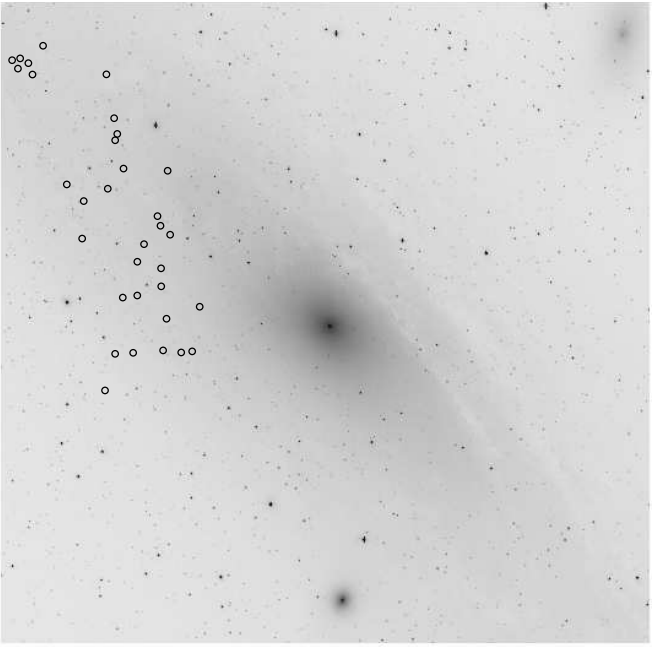}}}
\vspace{0.0cm} \caption{The location of the clusters
ontop of M31 image.} \label{fig:three}
\end{figure}

\subsection{Relationship between Age and Metallicity}

Among the methods for age determination for GCs in
the Galaxy, the most popular is the fitting of theoretical
isochrones to a cluster's CMD, which requires the accurate
measurement of $\Delta V$ (the difference in luminosity between
the turnoff and horizontal branch), along with an estimate of
[Fe/H]. Chaboyer et al. (1996) obtained ages for the 43 globular
cluster using the 10 different $M_V(\rm RR)$ relations. They
plotted the age as a function of metallicity, and found clear
relationship with the most metal-poor clusters being the oldest,
assuming $M_V(\rm{RR})=0.20[\rm{Fe/H}]+0.98$. Figure 6 plots the
metallicity as a function of age for the sample globular clusters.
It is clear that a relationship between age and metallicity is not
found in this figure.

\begin{figure}
\resizebox{\hsize}{!}{\rotatebox{-90}{\includegraphics{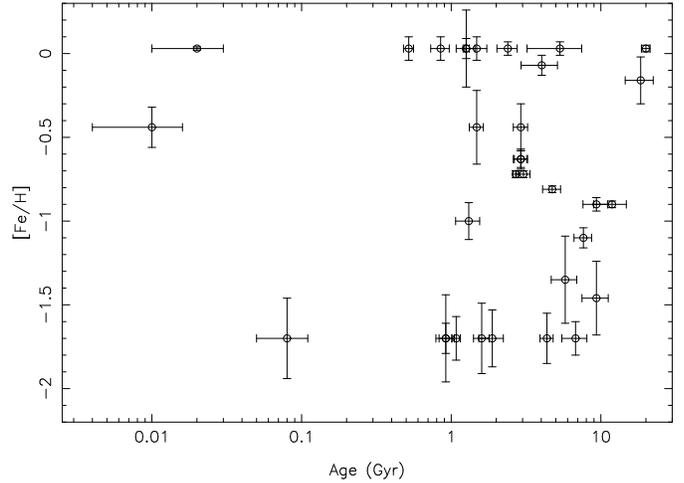}}}
\vspace{0.0cm} \caption{Metallicity as a function of age for
globular clusters of M31.} \label{fig:three}
\end{figure}

\section{Summary and Discussion}

We have obtained SEDs of 45 faint M31 GCs and GC candidates
detected by Mochejska et al. (1998) in 13 intermediate colors with
the NAOC 60/90 cm Schmidt telescope. By analyzing their SEDs, 3
are likely background objects. SED fits show that theoretical
simple stellar population (SSP) models can fit the true M31 GCs
very well. We confirm 20 candidates as M31 GCs, and we can
estimate their ages and metallicities by comparing with SSP
models. We find that these faint M31 GCs range in age from a few
ten Myr to a few Gyr old, as well as old GCs, confirming the
conclusions that has been found by Barmby et al. (2000),
Williams \& Hodge (2001), Beasley et al. (2004),
Burstein et al. (2004) and Puzia et al. (2005).

The fact that we can find faint GCs around M31 that have a similar
wide range of ages as the brighter GCs that Beasley et al. (2004) and
Burstein et al. (2004) have found is consistent with the formation
scenario proposed in Burstein et al. (2004). In that scenario, these
GCs come from dwarf galaxies that have been accreted by M31 in the
past. Evidently, the history of accretion of its contingent of dwarf
galaxies by M31 appears to be very different from that our Galaxy has
experienced.

In Table 5, there is a GC, the estimated age of which is 20 Gyr.
This age only means that this GC is as old as Galactic GCs.

\begin{acknowledgements}

We would like to thank the anonymous referee for his/her
insightful comments and suggestions that improved this paper very
much. This work has been supported by the Chinese National Key
Basic Research Science Foundation (NKBRSF TG199075402) and by the
Chinese National Natural Science Foundation, No. 10473012 and
10573020.
\end{acknowledgements}

\clearpage
\setcounter{table}{0}
\begin{table}[htb]
\caption[]{Parameters of the BATC Filters and Statistics of Observations for
M31}
\vspace {0.3cm}
\begin{tabular}{cccccc}
\hline
\hline
 No.& Name & $\rm {cw(\AA)}^{a}$ & Exp. (hr) & $\rm {N.img}^{b}$ & $\rm {rms}^{c}$ \\
\hline
        1  & BATC03 & 4210   & 01:00 & 03 & 0.015\\
        2  & BATC04 & 4546   & 05:30 & 17 & 0.009\\
        3  & BATC05 & 4872   & 03:30 & 11 & 0.015\\
        4  & BATC06 & 5250   & 02:20 & 12 & 0.006\\
        5  & BATC07 & 5785   & 02:15 & 07 & 0.003\\
        6  & BATC08 & 6075   & 01:40 & 05 & 0.003\\
        7  & BATC09 & 6710   & 00:45 & 03 & 0.003\\
        8  & BATC10 & 7010   & 03:00 & 12 & 0.008\\
        9  & BATC11 & 7530   & 02:00 & 06 & 0.004\\
       10  & BATC12 & 8000   & 04:00 & 12 & 0.003\\
       11  & BATC13 & 8510   & 01:30 & 05 & 0.004\\
       12  & BATC14 & 9170   & 05:50 & 18 & 0.003\\
       13  & BATC15 & 9720   & 04:00 & 12 & 0.009\\
\hline
\vspace{0.1cm}
\end{tabular}\\
$^{\rm a}~$ Central wavelength for each BATC filter\\
$^{\rm b}~$ Image numbers for each BATC filter\\
$^{\rm c}~$ Calibration error, in magnitude, for each filter as obtained from the standard stars\\
\end{table}

\clearpage
\setcounter{table}{1}
\begin{table}[htb]
\caption[]{SEDs of 45 GCs and GC Candidates in M31} \vspace
{0.5cm}
\begin{tabular}{cccccccccccccc}
\hline
\hline
$\rm {Cluster}$  & 03  &  04 &  05 &  06 &  07 &  08 &  09 &  10 &  11 &  12 &  13 &  14 &  15\\
(1)    & (2) & (3) & (4) & (5) & (6) & (7) & (8) & (9) & (10) & (11) & (12) & (13) & (14)\\
\hline
       M001 &  19.26 &  19.14 &  19.04 &  19.00 &  18.85 &  18.90 &  18.71 &  18.75 &  18.70 &  18.54 &  18.23 &  18.48 &  18.18\\
  &   0.118 &  0.138 &  0.163 &  0.162 &  0.193 &  0.207 &  0.200 &  0.236 &  0.272 &  0.272 &  0.252 &  0.339 &  0.301\\
       M002 &  18.53 &  18.41 &  18.22 &  18.17 &  18.03 &  18.09 &  17.94 &  17.89 &  18.07 &  17.80 &  17.98 &  17.81 &  18.14\\
  &   0.067 &  0.066 &  0.068 &  0.072 &  0.082 &  0.084 &  0.094 &  0.097 &  0.139 &  0.115 &  0.191 &  0.156 &  0.280\\
       M003 &  18.36 &  18.27 &  18.18 &  18.11 &  17.86 &  17.81 &  17.69 &  17.53 &  17.46 &  17.31 &  17.23 &  17.21 &  17.23\\
  &   0.078 &  0.073 &  0.073 &  0.075 &  0.069 &  0.064 &  0.072 &  0.063 &  0.068 &  0.059 &  0.074 &  0.080 &  0.117\\
       M005 &  19.79 &  19.79 &  19.71 &  19.70 &  19.59 &  19.55 &  19.67 &  19.50 &  19.40 &  19.70 &  19.84 &  18.52 &  18.54\\
  &   0.115 &  0.153 &  0.159 &  0.179 &  0.244 &  0.231 &  0.347 &  0.357 &  0.456 &  0.646 &  1.152 &  0.317 &  0.410\\
       M006 &  16.32 &  15.94 &  15.67 &  15.52 &  15.23 &  15.24 &  15.01 &  14.90 &  14.78 &  14.69 &  14.63 &  14.53 &  14.52\\
  &   0.007 &  0.005 &  0.005 &  0.005 &  0.005 &  0.005 &  0.006 &  0.005 &  0.006 &  0.006 &  0.009 &  0.008 &  0.011\\
       M007 &  15.92 &  15.54 &  15.25 &  15.13 &  14.82 &  14.87 &  14.86 &  14.50 &  14.42 &  14.32 &  14.29 &  14.20 &  14.20\\*
  &   0.005 &  0.004 &  0.004 &  0.003 &  0.004 &  0.004 &  0.005 &  0.004 &  0.005 &  0.004 &  0.007 &  0.007 &  0.008\\
       M009 &  18.43 &  18.14 &  17.83 &  17.69 &  17.46 &  17.40 &  17.24 &  17.21 &  17.11 &  17.00 &  17.00 &  16.96 &  16.95\\
  &   0.055 &  0.046 &  0.039 &  0.041 &  0.041 &  0.042 &  0.043 &  0.045 &  0.048 &  0.047 &  0.069 &  0.065 &  0.091\\
       M010 &  17.48 &  17.11 &  16.78 &  16.68 &  16.43 &  16.32 &  16.15 &  16.08 &  15.97 &  15.88 &  15.93 &  15.80 &  15.78\\
  &   0.012 &  0.009 &  0.011 &  0.010 &  0.013 &  0.011 &  0.013 &  0.014 &  0.017 &  0.017 &  0.029 &  0.023 &  0.033\\
       M011 &  17.92 &  17.92 &  17.66 &  17.58 &  17.37 &  17.33 &  17.21 &  17.08 &  16.99 &  16.94 &  16.83 &  16.78 &  16.87\\
  &   0.022 &  0.023 &  0.020 &  0.022 &  0.023 &  0.022 &  0.028 &  0.027 &  0.033 &  0.030 &  0.048 &  0.041 &  0.067\\
       M012 &  19.38 &  19.13 &  18.76 &  18.53 &  18.22 &  18.14 &  17.96 &  17.80 &  17.58 &  17.42 &  17.67 &  17.14 &  16.95\\
  &   0.071 &  0.071 &  0.061 &  0.055 &  0.064 &  0.059 &  0.060 &  0.065 &  0.071 &  0.060 &  0.131 &  0.074 &  0.077\\
       M013 &  17.30 &  17.06 &  16.79 &  16.67 &  16.39 &  16.38 &  16.23 &  16.13 &  16.06 &  15.95 &  15.89 &  15.86 &  15.87\\
  &   0.015 &  0.012 &  0.011 &  0.011 &  0.012 &  0.011 &  0.013 &  0.013 &  0.016 &  0.015 &  0.024 &  0.018 &  0.031\\
       M015 &  18.32 &  18.11 &  17.84 &  17.79 &  17.55 &  17.44 &  17.26 &  17.25 &  17.05 &  17.00 &  16.88 &  16.93 &  17.06\\
  &   0.037 &  0.037 &  0.038 &  0.038 &  0.043 &  0.038 &  0.037 &  0.043 &  0.048 &  0.047 &  0.064 &  0.067 &  0.097\\
       M016 &  19.80 &  19.71 &  19.45 &  19.37 &  19.49 &  19.31 &  19.09 &  19.28 &  18.99 &  19.49 &  19.63 &  18.34 &  18.40\\
  &   0.152 &  0.171 &  0.158 &  0.155 &  0.237 &  0.192 &  0.198 &  0.250 &  0.234 &  0.402 &  0.600 &  0.198 &  0.318\\
       M017 &  17.52 &  17.07 &  16.74 &  16.58 &  16.26 &  16.20 &  16.02 &  15.90 &  15.78 &  15.66 &  15.64 &  15.55 &  15.55\\
  &   0.014 &  0.009 &  0.007 &  0.008 &  0.010 &  0.008 &  0.009 &  0.010 &  0.011 &  0.009 &  0.015 &  0.016 &  0.025\\
       M018 &  19.25 &  18.65 &  18.01 &  17.62 &  17.30 &  17.09 &  16.77 &  16.69 &  16.44 &  16.29 &  16.07 &  15.90 &  15.88\\
  &   0.057 &  0.047 &  0.034 &  0.026 &  0.030 &  0.022 &  0.021 &  0.023 &  0.024 &  0.022 &  0.024 &  0.023 &  0.031\\
       M020 &  18.66 &  18.75 &  18.65 &  18.42 &  18.76 &  18.82 &  18.89 &  18.62 &  18.82 &  18.66 &  18.62 &  18.66 &  18.67\\
  &   0.091 &  0.099 &  0.098 &  0.080 &  0.127 &  0.116 &  0.153 &  0.131 &  0.191 &  0.166 &  0.248 &  0.259 &  0.372\\
         M021 &  15.67 &  15.28 &  14.99 &  14.83 &  14.54 &  14.49 &  14.30 &  14.20 &  14.09 &  13.96 &  13.93 &  13.85 &  13.82\\*
  &   0.005 &  0.003 &  0.003 &  0.003 &  0.003 &  0.003 &  0.003 &  0.003 &  0.003 &  0.003 &  0.004 &  0.004 &  0.005\\
       M022 &  19.52 &  19.20 &  18.88 &  18.83 &  18.51 &  18.60 &  18.43 &  18.23 &  18.11 &  18.09 &  18.33 &  17.96 &  17.76\\
  &   0.066 &  0.054 &  0.044 &  0.050 &  0.057 &  0.056 &  0.067 &  0.066 &  0.071 &  0.071 &  0.142 &  0.098 &  0.142\\
       M023 &  18.38 &  18.33 &  18.18 &  18.22 &  18.15 &  18.21 &  18.38 &  18.27 &  18.38 &  18.44 &  18.42 &  19.31 &  20.20\\
  &   0.077 &  0.068 &  0.070 &  0.078 &  0.091 &  0.096 &  0.143 &  0.125 &  0.181 &  0.192 &  0.246 &  0.579 &  1.777\\
       M024 &  17.40 &  17.08 &  16.77 &  16.68 &  16.36 &  16.33 &  16.20 &  16.14 &  16.04 &  15.92 &  15.91 &  15.86 &  15.87\\
  &   0.015 &  0.013 &  0.012 &  0.012 &  0.013 &  0.012 &  0.013 &  0.013 &  0.016 &  0.014 &  0.019 &  0.022 &  0.030\\
       M025 &  18.78 &  18.72 &  18.52 &  18.69 &  18.50 &  18.40 &  18.78 &  18.81 &  19.25 &  19.20 &  20.22 &  20.35 &  20.20\\
  &   0.074 &  0.087 &  0.088 &  0.107 &  0.124 &  0.111 &  0.162 &  0.182 &  0.299 &  0.310 &  0.939 &  1.218 &  1.510\\
       M026 & ...  &  22.25 & ...  &  21.07 &  20.00 &  19.78 &  19.10 &  18.61 &  18.30 &  18.14 &  17.98 &  17.65 &  17.47\\
     & ...  &  1.399 & ...  &  0.692 &  0.352 &  0.294 &  0.189 &  0.150 &  0.153 &  0.142 &  0.154 &  0.134 &  0.144\\
       M027 &  19.74 &  19.44 &  19.19 &  18.98 &  18.82 &  18.67 &  18.54 &  18.29 &  18.10 &  18.06 &  18.07 &  17.74 &  17.88\\
  &   0.163 &  0.133 &  0.131 &  0.103 &  0.109 &  0.089 &  0.097 &  0.075 &  0.079 &  0.081 &  0.117 &  0.088 &  0.167\\
       M028 &  19.02 &  18.84 &  18.63 &  18.59 &  18.43 &  18.23 &  18.47 &  18.33 &  18.32 &  18.15 &  18.35 &  18.33 &  18.87\\*
  &   0.076 &  0.073 &  0.071 &  0.071 &  0.080 &  0.067 &  0.095 &  0.098 &  0.130 &  0.107 &  0.191 &  0.204 &  0.462\\
       M029 &  17.86 &  17.38 &  17.08 &  16.92 &  16.53 &  16.48 &  16.30 &  16.18 &  16.03 &  15.90 &  15.85 &  15.83 &  15.75\\
  &   0.016 &  0.012 &  0.011 &  0.010 &  0.010 &  0.009 &  0.010 &  0.011 &  0.014 &  0.011 &  0.019 &  0.019 &  0.025\\
       M030 &  19.19 &  19.18 &  19.15 &  18.96 &  18.97 &  18.92 &  19.06 &  18.89 &  18.70 &  18.51 &  18.51 &  18.46 &  17.86\\
  &   0.099 &  0.128 &  0.147 &  0.132 &  0.173 &  0.167 &  0.209 &  0.200 &  0.228 &  0.203 &  0.269 &  0.254 &  0.196\\
\hline
\end{tabular}
\end{table}

\clearpage
\setcounter{table}{1}
\begin{table}[ht]
\caption[]{Continued}
\vspace {0.5cm}
\begin{tabular}{cccccccccccccc}
\hline
\hline
$\rm {Cluster}$  & 03  &  04 &  05 &  06 &  07 &  08 &  09 &  10 &  11 &  12 &  13 &  14 &  15\\
(1)    & (2) & (3) & (4) & (5) & (6) & (7) & (8) & (9) & (10) & (11) & (12) & (13) & (14)\\
\hline
       M032 &  18.20 &  17.82 &  17.64 &  17.50 &  17.28 &  17.26 &  17.08 &  17.00 &  16.86 &  16.82 &  16.77 &  16.58 &  16.51\\
  &   0.021 &  0.014 &  0.013 &  0.013 &  0.016 &  0.015 &  0.016 &  0.018 &  0.020 &  0.017 &  0.032 &  0.024 &  0.044\\
       M033 &  18.73 &  18.62 &  18.52 &  18.36 &  18.16 &  18.27 &  18.09 &  18.00 &  17.85 &  17.76 &  17.58 &  17.70 &  17.76\\
  &   0.073 &  0.074 &  0.083 &  0.077 &  0.083 &  0.089 &  0.082 &  0.081 &  0.091 &  0.090 &  0.109 &  0.105 &  0.156\\
       M035 &  16.22 &  15.92 &  15.67 &  15.54 &  15.28 &  15.24 &  15.10 &  15.02 &  14.92 &  14.83 &  14.80 &  14.77 &  14.76\\
  &   0.005 &  0.004 &  0.003 &  0.003 &  0.004 &  0.003 &  0.004 &  0.004 &  0.005 &  0.004 &  0.007 &  0.007 &  0.010\\
       M036 &  18.40 &  17.92 &  17.68 &  17.55 &  17.26 &  17.21 &  17.09 &  17.04 & ...  & ...  &  16.99 & ...  &  16.82\\*
  &   0.023 &  0.017 &  0.010 &  0.012 &  0.014 &  0.012 &  0.015 &  0.015 & ...  & ...  &  0.047 & ...  &  0.054\\
       M037 &  15.17 &  14.69 &  14.40 &  14.24 &  13.89 &  13.83 &  12.80 &  13.53 &  13.36 &  13.26 &  13.22 &  13.12 &  13.06\\
  &   0.004 &  0.003 &  0.003 &  0.002 &  0.002 &  0.002 &  0.001 &  0.002 &  0.002 &  0.002 &  0.003 &  0.002 &  0.003\\
       M038 &  19.13 &  18.51 &  18.10 &  17.76 &  17.55 &  17.44 &  17.15 &  17.07 & ... & ...  &  16.62 & ...  &  16.48\\
  &   0.038 &  0.028 &  0.013 &  0.013 &  0.017 &  0.012 &  0.015 &  0.013 & ...  & ...  &  0.038 & ...  &  0.044\\
       M043 &  20.57 &  20.25 &  20.03 &  19.94 &  19.61 &  19.70 &  19.66 &  19.83 &  19.53 &  19.24 &  19.16 &  18.84 &  20.21\\*
  &   0.329 &  0.252 &  0.225 &  0.212 &  0.213 &  0.218 &  0.277 &  0.344 &  0.359 &  0.291 &  0.411 &  0.298 &  1.696\\
       M045 &  19.82 &  19.51 &  19.21 &  18.95 &  18.71 &  18.50 &  18.26 &  18.08 &  17.75 &  17.47 &  17.68 &  17.44 &  17.21\\
  &   0.087 &  0.067 &  0.060 &  0.049 &  0.067 &  0.048 &  0.058 &  0.056 &  0.058 &  0.049 &  0.083 &  0.078 &  0.088\\
       M047 &  19.81 &  19.35 &  19.04 &  18.89 &  18.67 &  18.57 &  18.29 &  18.50 &  18.53 &  18.45 &  18.55 &  18.62 &  18.92\\
  &   0.172 &  0.090 &  0.089 &  0.090 &  0.096 &  0.084 &  0.076 &  0.117 &  0.153 &  0.146 &  0.186 &  0.253 &  0.506\\
       M048 &  18.14 &  17.76 &  17.45 &  17.36 &  16.98 &  16.94 &  16.79 &  16.70 &  16.58 &  16.46 &  16.44 &  16.39 &  16.35\\
  &   0.033 &  0.023 &  0.019 &  0.020 &  0.021 &  0.018 &  0.020 &  0.020 &  0.025 &  0.025 &  0.041 &  0.037 &  0.046\\
       M049 &  18.40 &  18.41 &  18.25 &  18.15 &  17.99 &  18.03 &  17.79 &  17.80 &  17.79 &  17.66 &  17.89 &  17.57 &  17.89\\
  &   0.049 &  0.055 &  0.056 &  0.050 &  0.054 &  0.054 &  0.066 &  0.062 &  0.070 &  0.066 &  0.102 &  0.090 &  0.198\\
       M051 &  17.67 &  17.33 &  17.03 &  16.86 &  16.56 &  16.50 &  16.25 &  16.23 &  16.07 &  15.92 &  15.94 &  15.87 &  15.78\\*
  &   0.046 &  0.030 &  0.027 &  0.020 &  0.019 &  0.018 &  0.020 &  0.015 &  0.015 &  0.015 &  0.017 &  0.019 &  0.029\\
       M052 &  20.16 &  20.02 &  19.95 &  19.94 &  19.69 &  19.49 &  19.45 &  19.50 &  19.11 &  19.15 &  19.05 &  18.92 &  18.36\\
  &   0.086 &  0.122 &  0.120 &  0.101 &  0.144 &  0.098 &  0.133 &  0.167 &  0.206 &  0.210 &  0.356 &  0.287 &  0.256\\
       M056 &  20.50 &  19.91 &  19.75 &  19.71 &  19.51 &  19.35 &  19.28 &  19.20 &  18.86 &  18.89 &  18.35 &  18.57 &  18.52\\
  &   0.236 &  0.167 &  0.146 &  0.131 &  0.216 &  0.149 &  0.198 &  0.189 &  0.228 &  0.261 &  0.258 &  0.316 &  0.357\\
       M058 &  20.10 &  19.64 &  19.17 &  18.97 &  18.43 &  18.38 &  17.99 &  17.90 &  17.58 &  17.42 &  17.32 &  17.41 &  17.29\\
  &   0.160 &  0.118 &  0.087 &  0.071 &  0.064 &  0.057 &  0.049 &  0.053 &  0.059 &  0.055 &  0.078 &  0.086 &  0.109\\
       M061 &  19.54 &  18.97 &  18.61 &  18.52 &  18.15 &  18.12 &  17.96 &  17.75 &  17.58 &  17.49 &  17.54 &  17.33 & ... \\
  &   0.068 &  0.042 &  0.039 &  0.037 &  0.039 &  0.038 &  0.035 &  0.035 &  0.045 &  0.043 &  0.069 &  0.062 & ... \\
       M063 &  18.44 &  18.34 &  18.23 &  18.27 &  18.09 &  18.13 &  17.99 &  18.25 &  18.08 &  18.06 &  18.47 &  18.59 & ... \\*
  &   0.078 &  0.077 &  0.086 &  0.083 &  0.091 &  0.092 &  0.080 &  0.115 &  0.119 &  0.137 &  0.248 &  0.276 & ... \\
       M065 &  19.41 &  19.21 &  18.96 &  19.09 &  18.84 &  18.91 &  18.92 &  18.82 &  19.06 & ...  &  19.16 &  20.13 & ... \\
  &   0.065 &  0.055 &  0.058 &  0.060 &  0.075 &  0.078 &  0.104 &  0.122 &  0.220 & ...  &  0.389 &  0.884 & ... \\
       M066 &  18.14 &  17.72 &  17.33 &  17.22 &  16.82 &  16.74 &  16.64 &  16.55 &  16.41 & ...  &  16.31 &  16.28 & ... \\
  &   0.034 &  0.023 &  0.019 &  0.018 &  0.019 &  0.015 &  0.013 &  0.016 &  0.020 & ...  &  0.024 &  0.026 & ... \\

\hline
\end{tabular}
\end{table}

\clearpage
\setcounter{table}{2}
\begin{table}[htb]
\caption[]{Results of SED Fits between SSP Models and GCs} \vspace
{0.3cm}
\begin{tabular}{cccc}
\hline
\hline
    Name  & Age  & Metallicity & $\chi^2_{min}$\\
          &(Gyr) & ([Fe/H])   & per degree of freedom \\
\hline
M006~~~  Bo205   & $  4.73 \pm  0.65 $ & $ -0.81 \pm  0.02 $ &  0.47  \\
M007~~~  Bo206   & $  1.48 \pm  0.25 $ & $  0.00 \pm  0.07 $ &  2.64  \\
M009~~~  Bo209   & $  2.92 \pm  0.32 $ & $ -0.63 \pm  0.06 $ &  0.64  \\
M011~~~  Bo210   & $  1.88 \pm  0.35 $ & $ -1.70 \pm  0.17 $ &  1.19  \\
M013~~~  Bo211   & $  7.65 \pm  1.03 $ & $ -1.10 \pm  0.06 $ &  0.40  \\
M015~~~  Bo214   & $  5.78 \pm  1.12 $ & $ -1.35 \pm  0.26 $ &  0.63  \\
M017~~~  Bo217   & $  9.34 \pm  1.74 $ & $ -0.90 \pm  0.04 $ &  0.75  \\
M021~~~  Bo218   & $  2.70 \pm  0.12 $ & $ -0.72 \pm  0.02 $ &  0.31  \\
M024~~~  Bo220   & $  9.34 \pm  1.86 $ & $ -1.46 \pm  0.22 $ &  0.66  \\
M029~~~  Bo221   & $  3.04 \pm  0.32 $ & $ -0.72 \pm  0.02 $ &  0.56  \\
M032~~~  Bo222   & $  1.26 \pm  0.06 $ & $  0.00 \pm  0.06 $ &  0.45  \\
M035~~~  Bo224   & $  4.36 \pm  0.44 $ & $ -1.70 \pm  0.15 $ &  0.37  \\
M048~~~  Bo231   & $  2.92 \pm  0.28 $ & $ -0.63 \pm  0.05 $ &  0.43  \\
M051~~~  Bo234   & $ 11.88 \pm  2.92 $ & $ -0.90 \pm  0.02 $ &  0.36  \\
\hline
\end{tabular}
\end{table}

\clearpage
\setcounter{table}{3}
\begin{table}[htb]
\caption[]{Comparison of Results}
\vspace {0.3cm}
\begin{tabular}{cccc|ccc}
\hline
\hline
    Name & Age & Metallicity & $\chi^2_{min}$          &
    Age  & Metallicity& $\chi^2_{min}$\\
         & (Gyr)     &   ([Fe/H]) & per degree of freedom &
    (Gyr)& ([Fe/H])  & per degree of freedom \\
\hline
M009~~~  Bo209   & $  2.59 \pm  0.13 $ & $ -0.72 \pm  0.04 $ &  0.74     & $  2.92 \pm  0.32 $ & $ -0.63 \pm  0.06 $ &  0.64  \\
M013~~~  Bo211   & $  4.92 \pm  0.57 $ & $ -1.70 \pm  0.15 $ &  0.35     & $  7.65 \pm  1.03 $ & $ -1.10 \pm  0.06 $ &  0.40  \\
M015~~~  Bo214   & $  5.12 \pm  0.92 $ & $ -1.70 \pm  0.12 $ &  0.60     & $  5.78 \pm  1.12 $ & $ -1.35 \pm  0.26 $ &  0.63  \\
M017~~~  Bo217   & $  6.78 \pm  0.79 $ & $ -0.72 \pm  0.02 $ &  0.70     & $  9.34 \pm  1.74 $ & $ -0.90 \pm  0.04 $ &  0.75  \\
M021~~~  Bo218   & $  3.17 \pm  0.43 $ & $ -0.63 \pm  0.02 $ &  0.30     & $  2.70 \pm  0.12 $ & $ -0.72 \pm  0.02 $ &  0.31  \\
M024~~~  Bo220   & $  7.06 \pm  1.28 $ & $ -1.70 \pm  0.15 $ &  0.74     & $  9.34 \pm  1.86 $ & $ -1.46 \pm  0.22 $ &  0.66  \\
\hline
\end{tabular}
\end{table}

\clearpage
\setcounter{table}{4}
\begin{table}[htb]
\caption[]{Results of SED Fits between SSP Models and GC
Candidates} \vspace {0.3cm}
\begin{tabular}{cccc}
\hline
\hline
    Name  & Age  & Metallicity & $\chi^2_{min}$\\
          &(Gyr) & ([Fe/H])   & per degree of freedom \\
\hline
 M001    & $  0.52 \pm  0.04 $ & $  0.03 \pm  0.07 $ &  0.20  \\
 M002    & $  0.92 \pm  0.09 $ & $ -1.70 \pm  0.09 $ &  0.60  \\
 M003    & $  1.31 \pm  0.24 $ & $ -1.00 \pm  0.11 $ &  0.91  \\
 M009    & $  6.78 \pm  1.29 $ & $ -1.70 \pm  0.10 $ &  0.44  \\
 M012    & $  5.33 \pm  2.12 $ & $  0.03 \pm  0.04 $ &  1.49  \\
 M018      &      ...   &      ...   & 10.96  \\
 M020    & $  0.08 \pm  0.03 $ & $ -1.70 \pm  0.24 $ &  1.06  \\
 M022    & $  1.48 \pm  0.16 $ & $ -0.44 \pm  0.22 $ &  1.06  \\
 M023      &      ...   &      ...   &  3.25  \\
 M025      &      ...   &      ...   &  5.32  \\
 M027    & $  2.39 \pm  0.37 $ & $  0.03 \pm  0.04 $ &  0.57  \\
 M028    & $  0.92 \pm  0.13 $ & $ -1.70 \pm  0.26 $ &  1.66  \\
 M030    & $  0.02 \pm  0.00 $ & $  0.03 \pm  0.01 $ &  0.68  \\
 M033    & $  1.60 \pm  0.19 $ & $ -1.70 \pm  0.21 $ &  0.64  \\
 M036      &      ...   &      ...   &  3.05  \\
 M038    & $ 18.46 \pm  3.92 $ & $ -0.16 \pm  0.14 $ &  2.35  \\
 M045      &      ...   &      ...   &  3.19  \\
 M047      &      ...   &      ...   &  3.24  \\
 M049    & $  1.08 \pm  0.07 $ & $ -1.70 \pm  0.13 $ &  1.11  \\
 M052    & $  0.85 \pm  0.12 $ & $  0.03 \pm  0.07 $ &  1.21  \\
 M056    & $  1.26 \pm  0.18 $ & $  0.03 \pm  0.23 $ &  0.44  \\
 M058    & $ 20.00 \pm  1.34 $ & $  0.03 \pm  0.02 $ &  2.16  \\
 M061    & $  4.03 \pm  1.10 $ & $ -0.07 \pm  0.06 $ &  0.74  \\
 M063    & $  0.01 \pm  0.00 $ & $ -0.44 \pm  0.12 $ &  1.49  \\
 M065      &      ...   &      ...   &  4.24  \\
 M066    & $  2.92 \pm  0.32 $ & $ -0.44 \pm  0.14 $ &  1.72  \\
\hline
\end{tabular}
\end{table}

\end{document}